\documentstyle[twoside,fleqn,espcrc2,epsf]{article}
\newcommand{\ewxy}[2]{\setlength{\epsfxsize}{#2}\epsfbox[10 60 640 590]{#1}}
\def\be{\begin{equation}}
 
\def\ee{\end{equation}}
 
\def\bea{\begin{eqnarray}}
 
\def\eea{\end{eqnarray}}
 
\newcommand{\gev}{\mbox{$\;$GeV}}

\newcommand{\PP}{I\!\!P}

\input psfig 

\newcommand{\AmS}{{\protect\the\textfont2
  A\kern-.1667em\lower.5ex\hbox{M}\kern-.125emS}}

\title{Pomerons and Lattices: a Progress Report}

\author{
{\it UKQCD Collaboration} - presented by C. 
Parrinello\address{Department of Mathematical Sciences, 
University of Liverpool,\\
Liverpool L69 3BX, United Kingdom}}

\begin{document}

\begin{abstract}
We report on some attempts to use  
 lattice QCD to investigate
 topics in strong interaction phenomenology which
are usually interpreted in terms of soft Pomeron exchange.
\end{abstract}

\maketitle

\section{INTRODUCTION}

It has been known for a long time
 that the rise of the total hadronic 
cross section at high energies could be described in 
the framework of Regge theory via Pomeron exchange 
\cite{land_DIS}. 
Renewed 
interest in Pomeron 
models has been recently triggered by the analysis of diffractive 
scattering at HERA. In particular, the discovery of events with large rapidity 
gaps has been interpreted as evidence for Pomeron 
exchange \cite{HERA1}.
Recovering the 
phenomenology of Pomeron exchange from first principles QCD 
is a major theoretical problem. 
 Here we describe some attempts to use lattice QCD 
 to test models of diffraction and the Pomeron.
We first describe an application of the method to  
 the Landshoff-Nachtmann (LN) model \cite{LN} and then 
 we outline a more general approach to the 
lattice investigation of diffractive physics. 

\section{THE LN POMERON} 

In the LN model 
 Pomeron exchange between quarks behaves like a
$C=+1$ photon-exchange
diagram, with amplitude
\begin{equation}
i \beta^2_0 (\bar{u} \gamma_\mu u)(\bar{u} \gamma^\mu u).
\end{equation}
$\beta_0$ represents the strength of the Pomeron coupling to quarks,
and is related to the (non-perturbative) gluon propagator $D(p)$ by
\begin{equation}
\beta_0^2 = \frac{1}{36 \pi^2} \int d^2 p \left[ g^2 D(p)\right]^2,
\label{eq:beta0_constraint}
\end{equation}
where $g$ is the gluon-quark coupling.
The model yields simple formulae for $pp$ scattering,
exclusive $\rho$ production in deep inelastic scattering and the $J/ \Psi - $
nucleon total cross section, which all contain integrals in momentum space of
$g^2 D(p)$ \cite{ducati:93}.

In order to make predictions one needs an expression
for $g^2 D(p)$. 
By inserting in the model the gluon propagator
 as computed on the lattice (in the Landau gauge) we aimed to perform 
a strong consistency test 
from the point of view of QCD \cite{noi}. 
For this purpose we generated two sets of quenched configurations 
and in 
addition we used data
for the gluon propagator as evaluated in \protect\cite{stella:94}.
The lattice parameters are
listed in Table~\ref{tab:params}.  Full details of the results and 
the method for the determination of the 
gluon propagator 
can be found in ref. \cite{noi}. In the figures the results obtained from 
the data sets 1, 2 and 3 are plotted with a dotted, solid
 and dashed line respectively.
\begin{table*}[hbt]
\setlength{\tabcolsep}{1.5pc}
\newlength{\digitwidth} \settowidth{\digitwidth}{\rm 0}
\catcode`?=\active \def?{\kern\digitwidth}\caption{Parameters of the 
lattices used in the study of the LN model.}\label{tab:params}
\begin{tabular*}{\textwidth}{@{}l@{\extracolsep{\fill}}rrrr}
\hline
                 & \multicolumn{1}{r}{$\beta$}
                 & \multicolumn{1}{r}{Size}
                 & \multicolumn{1}{r}{Cfgs.}
                 & \multicolumn{1}{r}{$a^{-1}\ (\gev)$} \\
\hline
Data set 1 & 6.0 & $16^4$ & 150 &  1.9 \\ 
Data set 2 & 6.2 & $16^4$ & 150 &  2.7 \\ 
Data set 3 \protect\cite{stella:94} & 6.0 & $24^3 \times 48$ & 500 & 1.9 \\ 
\hline
\end{tabular*}
\end{table*}
We inserted in the LN formulae analytical expressions
 corresponding to the best 
fits to the lattice gluon propagator on each data set.
It turns out that our calculations are not significantly 
affected by 
discretisation effects, as all the relevant integrals in momentum space 
appeared to reach their asymptotic value at an integration
 scale well below the inverse 
lattice spacing. 
It was assumed that in the continuum limit
the propagator
is multiplicatively renormalisable, as it is in perturbation theory.
Also, we neglected the running of the QCD coupling, i.e.
we made the approximation $g(p) = g$.
As the scale for the momenta in $D (p)$ was set
from independent string tension measurements, we only had one
free parameter  to fix in the expression $g^2 D (p)$.
This is a multiplicative factor 
corresponding to the product of a gluon wavefunction renormalisation constant
times a numerical value for $g^2$.
We call this parameter $g_{\rm eff}^2$.
It was determined by using (\ref{eq:beta0_constraint}) as a
nonperturbative 
renormalisation condition, imposing that $\beta_0$ attains
its phenomenological value of 2.0 ${\gev}^{-1}$ \cite{landshoff:84}:
\begin{equation}
\beta_{0}^2 = \frac{1}{36 \pi^2} \int d^2 p \left[ g_{\rm eff}^2 \ 
D_{\rm lat} (p)
\right]^2 = 4 \ {\gev}^{-2}.
\label{eq:beta0_fix}
\end{equation}
We finally inserted $g_{\rm eff}^2 D_{\rm lat} (p)$ in the
formulae of the model. 
We adopted the analysis procedure of Halzen and 
collaborators \cite{ducati:93}.
 
\subsection{Proton-proton elastic scattering}
 
The calculation of $\sigma^0_{\rm tot}$ and $\frac{d \sigma^0}{dt}$, i.e. 
the energy-independent part of the 
total and elastic differential cross section for proton-proton
scattering, provides a benchmark for the LN model. 
Single Pomeron exchange is expected to dominate
$\frac{d \sigma}{dt}$ up to $-t \simeq 0.5 \, {\gev}^2$.  
The measured
total and elastic 
differential cross sections are usually parametrised as
follows:
\begin{equation}
\sigma_{\rm tot}  =  \left(\frac{s}{m_p^2}\right)^{0.08}
\sigma^0_{\rm tot}, \ 
\frac{d \sigma}{dt}  =  \left(\frac{s}{m_p^2}\right)^{0.168}
\frac{d \sigma^0}{dt}. 
\label{eq:dsig_energy}
\end{equation}
For small $t$, the elastic differential cross section behaves like
$e^{B t}$, and the model is characterised by two parameters,
$\sigma^0_{\rm tot}$ and $B$.
 
We computed
$\sigma^0_{\rm tot}$ and $B$ on each lattice using the lattice
gluon propagator and the effective coupling $g_{\rm eff}$.
We obtained values of $\sigma^0_{\rm tot}$ ranging from  $18.12 \ {\rm mb}$ to 
$19.85 \ {\rm mb}$ 
and
$B$ from $12.6 {\gev}^{-2}$ to $13.6 {\gev}^{-2}$, thus
in very good agreement with 
each other, suggesting 
that both quantities are subject to only small 
finite volume effects.
 They are also encouragingly close to the phenomenological
values of $\sigma^0_{\rm tot} \simeq 22.7\,{\rm mb}$ and $B \sim
11~\gev^{-2}$.
\begin{figure}[t]
\vspace{-5mm}
\ewxy{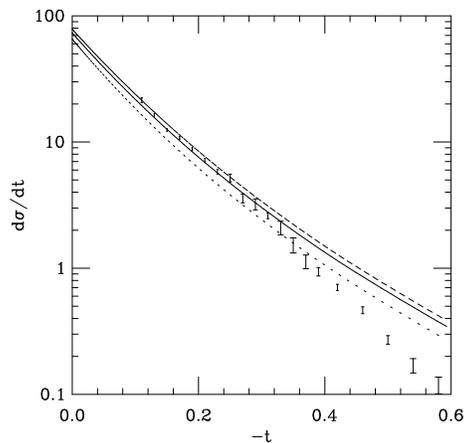}{80mm}
\caption{Data 
 for the $pp$ elastic cross section at $\protect\sqrt{s} =
53~\gev$ together with the lattice
predictions, corrected for the energy dependence.}
\label{fig:isr_data}
\end{figure}
In Fig.~\ref{fig:isr_data} we show ISR data \protect\cite{isr:84}
 for the differential
elastic cross section at $\sqrt{s} = 53~\gev$ together with the
lattice predictions, with the energy correction of
Eq.~\ref{eq:dsig_energy}. 
 
\subsection{$J/\psi$-nucleon scattering}

This process provides a further important test of the LN model.
\begin{figure}[t]
\vspace{-5mm}
\ewxy{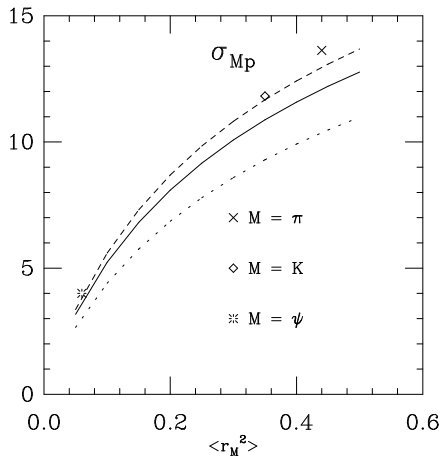}{80mm}
\caption{The meson-nucleon total cross section as a function of the
radius used in the meson form factor.
 Also shown
are the radii and total cross sections for the $\pi$, $K$
and $J/\psi$.}
\label{fig:jpsi}
\end{figure}
Two phenomenological features emerge from our calculations.  Firstly, the
quark-counting rule is closely satisfied for light mesons,
 as we get $\sigma_{pp}/\sigma_{\pi p}$ 
 in the range $1.5 - 1.8$.
Secondly, the Pomeron couples more weakly to heavier mesons, as
can be seen from
Fig.~\ref{fig:jpsi},
 where we show the meson-nucleon cross section as a function
of the pole radius, together with a Regge
fit to the energy-independent part of the $\pi^{-} p$ and $K^{-} p$
cross sections \cite{land_DIS}. 
Sizeable discrepancies
between different lattices can be observed in this case, which may be related 
to finite volume effects.
In this connection, it is worth noting that 
for most quantities the results obtained from the largest lattice provide
the best agreement with the experimental values.
In order to further investigate this point 
we plan to repeat our calculations for a wider range of lattice parameters.

\section{DIFFRACTION ON THE LATTICE}

Here we summarise 
a more general approach for lattice investigations of diffractive
scattering. 
We start by assuming a 
{\it factorisable} Pomeron exchange.  
To clarify this concept, consider diffractive scattering at HERA in the 
one-photon approximation:
\begin{equation}
\gamma^*(q)+p(p)\to\tilde p(\tilde p)+X(p_X).
\label{eq:scat}
\end{equation}

If the above process factorises, one can describe it in two steps 
\cite{landlast}:
first the original proton emits a Pomeron,  
$p(p) \rightarrow p(\tilde p)+ \PP(\Delta)$,
then the Pomeron collides with
the $\gamma^*$ to produce the final hadronic state $X$,
$\gamma^*(q)+ \PP (\Delta) \rightarrow X(p_X)$.
On the lattice we aim to study the first subprocess, calculating  
the effective coupling of the Pomeron to the proton. One motivation for such 
a study is the investigation of the helicity structure of the Pomeron, as it
 has 
recently been argued that a nontrivial spin structure in the 
Pomeron-proton vertex may affect quantities measured at 
HERA \cite{landlast}. 
 Given some {\it ansatz} for a composite operator 
$O_{\PP} (x)$ which creates the Pomeron from the QCD vacuum, the effective 
Pomeron-proton coupling  takes the form of a QCD 3-point vertex function, which we can study in momentum space as a function of the momentum of the proton, 
$p$, and the momentum of the Pomeron, $\Delta$.
We 
assume that $O_{\PP} $ only contains gluonic fields. 
As for the quantum numbers, since the Pomeron is a Regge trajectory
all the information we have {\it a priori} is that $O_{\PP}$ should have 
a $J^{++}$ structure.
As we do not want to assign a value to $J$, it makes sense to 
allow $O_{\PP}$ to have arbitrary Lorentz structure, i.e. to be the sum
of a scalar, a vector, and tensors of arbitrary rank. 

Numerical work is in progress.
We are currently focusing on two-gluon operators for $O_{\PP}$ with 
 encouraging preliminary results.

\section*{ACKNOWLEDGEMENTS}

We acknowledge support from 
PPARC under
grant GR/J21347 and 
Advanced Fellowships (DGR and CP).

\end{document}